\documentclass[twocolumn,prl]{revtex4}
\usepackage{graphicx}
\usepackage{hyperref}
\usepackage{times}

\begin{document}

\title{The New Minimal Standard Model}

\author{Hooman Davoudiasl}
\affiliation{School of Natural Sciences, Institute for Advanced Study,
  Einstein Drive, Princeton, NJ 08540, USA}

\author{Ryuichiro Kitano}
\affiliation{School of Natural Sciences, Institute for Advanced Study,
  Einstein Drive, Princeton, NJ 08540, USA}

\author{Tianjun Li}
\affiliation{School of Natural Sciences, Institute for Advanced Study,
  Einstein Drive, Princeton, NJ 08540, USA}

\author{Hitoshi Murayama} \thanks{On leave of absence from Department
  of Physics, University of California, Berkeley, CA 94720.}
\affiliation{School of Natural Sciences, Institute for Advanced Study,
  Einstein Drive, Princeton, NJ 08540, USA}

\date{May 11, 2004}

%%%%%%%%%%%%%%%%%%%%%%%%%%%%%%%%%%%%%%%%%%%%%%%%%%%%%%%%%%%%%%%%%%%%%%%%%%%%

\begin{abstract}
  We construct the New Minimal Standard Model that incorporates the
  new discoveries of physics beyond the Minimal Standard Model (MSM):
  Dark Energy, non-baryonic Dark Matter, neutrino masses, as well as
  baryon asymmetry and cosmic inflation, adopting the principle of
  minimal particle content and the most general renormalizable
  Lagrangian.  We base the model purely on empirical facts rather than
  aesthetics.  We need only six new degrees of freedom beyond the MSM.
  It is free from excessive flavor-changing effects, CP violation,
  too-rapid proton decay, problems with electroweak precision data,
  and unwanted cosmological relics.  Any model of physics beyond the
  MSM should be measured against the phenomenological success of this
  model.
\end{abstract}
\maketitle

The last several years have brought us revolutionary new insights into
fundamental physics: the discovery of Dark Energy, neutrino masses and
bi-large mixings, a solid case for non-baryonic Dark Matter, and
mounting evidence for cosmic inflation. It is now clear that the
age-tested Minimal Standard Model (MSM) is incomplete and needs to be
expanded.

There exist many possible directions to go beyond the MSM:
supersymmetry, extra dimensions, extra gauge symmetries ({\it
e.g.}\/, grand unification), etc. They are motivated to solve
aesthetic and theoretical problems of the MSM, but not necessarily
to address empirical problems.  It is embarrassing that all
currently proposed frameworks have some phenomenological problems,
{\it e.g.}\/, excessive flavor-changing effects, CP violation,
too-rapid proton decay, disagreement with electroweak precision
data, and unwanted cosmological relics.

In this letter, we advocate a different and conservative approach
to physics beyond the MSM.  We include the minimal number of new
degrees of freedom to accommodate convincing ({\it e.g.}\/, $>
5\sigma$) evidence for physics beyond the MSM.  We do not pay
attention to aesthetic problems, such as fine-tuning, the
hierarchy problem, etc. We stick to the principle of minimality
seriously to write down the Lagrangian that explains everything we
know.  We call such a model the New Minimal Standard Model (NMSM).
In fact, the MSM itself had been constructed in this spirit, and
it is a useful exercise to follow through with the same logic at
the advent of the major discoveries we have witnessed.  Of course,
we require it to be a consistent Lorentz-invariant renormalizable
four-dimensional quantum field theory, the way the MSM was
constructed.

We should not forget that the MSM is a tremendous success of the
twentieth century physics.  It is a gauge theory based on the $SU(3)_C
\times SU(2)_L \times U(1)_Y$ gauge group, has three generations of
quarks and leptons, one doublet Higgs boson, and a completely general
renormalizable Lagrangian one can write down.  We also add {\it
  classical}\/ gravity for completeness.  The Lagrangian can be
written down in a few lines (we omit the metric factor $\sqrt{-g}$):
\begin{eqnarray}
  \label{eq:MSM}
  \lefteqn{
    {\cal L}_{MSM} =
    -\frac{1}{2 g_s^2} {\rm Tr} G_{\mu\nu} G^{\mu\nu}
    - \frac{1}{2 g^2} {\rm Tr} W_{\mu\nu} W^{\mu\nu} } \nonumber \\
    & & - \frac{1}{4 g^{\prime 2}} B_{\mu\nu} B^{\mu\nu}
  + i\frac{\theta}{16\pi^2} {\rm Tr} G_{\mu\nu} \tilde{G}^{\mu\nu}
  + M_{Pl}^2 R
  \nonumber \\
  & & + |D_\mu H|^2 + \bar{Q}_i i{\not\!\!D} Q_i
  + \bar{U}_i i{\not\!\!D} U_i + \bar{D}_i i{\not\!\!D} D_i
  \nonumber \\
  & & + \bar{L}_i i{\not\!\!D} L_i+ \bar{E}_i i{\not\!\!D} E_i
  - \frac{\lambda}{2} \left(H^\dagger H - \frac{v^2}{2} \right)^2
  \nonumber \\
  & & - \left( h_u^{ij} Q_i U_j \tilde{H}
    + h_d^{ij} Q_i D_j H + h_l^{ij} L_i E_j H + c.c.\right).
\end{eqnarray}
Here, $M_{Pl} = 2.4 \times 10^{18}$~GeV is the reduced Planck
constant, $\tilde{H} = i\sigma_2 H^*$, and $i,j=1,2,3$ are generation
indices.  It is quite remarkable that the nineteen physically
independent parameters in these few lines explain nearly all phenomena
we have observed in our universe.

Using the principle of minimal particle content, we attempt to
construct the NMSM.  It is supposed to be the complete theory up to
the Planck scale unless experiments guide us otherwise.  What is such
a theory?  We claim we need only four new particles beyond the MSM to
construct the NMSM, two Majorana spinors and two real scalars, or six
degrees of freedom.  Note that all components we add to the MSM had
been used elsewhere in the literature.  What is {\it new}\/ in our
model is that (1) it is inclusive, namely it covers all the recent
important discoveries listed below, and (2) it is consistent, namely
that different pieces do not conflict with each other or with the
empirical constraints.  Even though the latter may not appear an
important point, it is worth recalling that incorporating two
attractive ideas often leads to tensions and/or conflict, {\it
  e.g.}\/, supersymmetry and electroweak baryogenesis because of the
constraints from the electric dipole moments, axion dark matter and
string theory because of the cosmological overabundance, leptogenesis
and supersymmetry because of the gravitino problem, etc.  We find it
remarkable and encouraging that none of the elements we add to the MSM
cause tensions nor conflicts which we will verify explicitly in the
letter.

What physics do we need to incorporate into the NMSM that is lacking
in the MSM?  Here is the list:

\noindent $\bullet$ Dark Matter has been suggested as a necessary ingredient of
cosmology for various reasons.  There is now compelling evidence for a
non-baryonic matter component \cite{Spergel:2003cb}.

\noindent $\bullet$ Dark Energy is needed based on the concordance of data from
cosmic microwave anisotropy \cite{Spergel:2003cb}, galaxy clusters
(see, {\it e.g.}\/, \cite{Verde:2001sf}), and high-redshift Type-IA
supernovae \cite{Perlmutter:1998np,Riess:1998cb}.

\noindent $\bullet$ Atmospheric \cite{Fukuda:1998mi} and solar
neutrino oscillations \cite{Ahmed:2003kj} have been established, with
additional support from reactor anti-neutrinos \cite{Eguchi:2003gg},
demonstrating neutrino masses and mixings.

\noindent $\bullet$ The cosmic baryon asymmetry $\eta = n_B/s =
9.2^{+0.6}_{-0.4} \times 10^{-11}$, which cannot be explained in the
MSM, has been known for many decades.

\noindent $\bullet$ The nearly scale-invariant, adiabatic, and
Gaussian density fluctuations (see, {\it e.g.}\/,
\cite{Komatsu:2003fd}) point to cosmic inflation.  This has not been
proven, but we find the evidence compelling.

%% There are many other empirical hints for physics beyond the MSM,
%% including NuTeV anomaly, LSND neutrino oscillation signature, the
%% apparent lack of cusps in galaxies, the apparent lack of rich
%% substructure in galaxy density profile, high hadronic cross
%% sections at $e^+ e^-$ colliders, CP violation in $B_d \rightarrow
%% \phi K_S$, muon $g-2$, excess $b$-jets at hadron colliders, etc.
%% Even though they are all interesting, we do not regard them as
%% convincing or urgent enough to merit inclusion into the NMSM. One
%% possible exception may be the claimed Dark Matter detection in the
%% annual modulation of the event rates in the DAMA detector, but we
%% only discuss Dark Matter on more generic terms, independent of
%% this controversial specific signal.
There are many other hints for physics beyond the MSM at a few sigma
levels which we do not try to incorporate.

We now apply our principle of minimal particle content to address
each of the issues.  First, we discuss Dark Matter.  It is clear
that the MSM does not have a candidate degree of freedom.  The
minimal way to add a new degree of freedom in a quantum field
theory is a real Klein--Gordon (KG) field.  To make it stable, we
must assign it a symmetry; the only such possibility for a real KG
field is a $Z_2$ parity.  Therefore, we introduce a singlet field
$S$ completely neutral under the gauge group and odd under a $Z_2$
parity. Then its most general renormalizable Lagrangian is
\begin{equation}
  \label{eq:S}
  {\cal L}_S =
  \frac{1}{2} \partial_\mu S \partial^\mu S - \frac{1}{2} m_S^2 S^2
  - \frac{k}{2} |H|^2 S^2 - \frac{h}{4!} S^4.
\end{equation}
It is encouraging that this model indeed had been proposed to explain
the cosmological Dark Matter in the past
\cite{Silveira:1985rk,McDonald:1993ex,Burgess:2000yq}.  Remarkably,
this model can explain the correct abundance, the lack of its
detection so far, and the lack of observation at high-energy
accelerators.  We will show later that the model is still viable.
This is clearly the minimal model of Dark Matter.

The next issue is Dark Energy.  Because we do not concern
ourselves with aesthetic issues such as naturalness and
fine-tuning in constructing the NMSM, we simply postulate a
cosmological constant of the observed size, approximately
\begin{equation}
  \label{eq:Lambda}
  {\cal L}_\Lambda = (2.3 \times 10^{-3}~{\rm eV})^4.
\end{equation}
This is a relevant operator in the Lagrangian, consistent with all
known symmetries.  Hence, it cannot be left out in a most general
Lagrangian.  Its renormalized value at the Hubble scale needs to be
the one given above.

The third issue is the neutrino masses and bi-large mixings.  We have
strong evidence for two mass-squared splittings, one from atmospheric
neutrinos $\Delta m^2 \simeq 2.5 \times 10^{-3}$~eV$^2$, and the other
from solar neutrinos (and reactor anti-neutrinos) $\Delta m^2 \simeq 7
\times 10^{-5}$~eV$^2$. Because the Planck-scale operator $(L{\tilde
  H})(L{\tilde H})/M_{Pl}$ gives only $m_\nu \lesssim 10^{-5}$~eV, too
small to explain the data, we need new degrees of freedom to
generate neutrino masses.  There is no evidence that all three
neutrinos are massive, and one of them may be exactly massless. We
hence need only two right-handed neutrinos $N_\alpha$ ($\alpha=1,2$),
or four new degrees of freedom, to write down the mass terms. We still
have to make a choice whether the mass terms are of Dirac or Majorana
type.  Based on the minimality alone, either of them is perfectly
valid. In the case of Dirac neutrinos, we need to impose a global
lepton number symmetry, while for Majorana neutrinos, we write down
all possible renormalizable terms.  The next minimal way of generating
Majorana neutrino masses requires a triplet scalar exchange
\cite{Lazarides:1980nt} with six new degrees of freedom.  Therefore,
adding two right-handed neutrinos is the minimal choice.

Next, we have to explain the baryon asymmetry of the universe.  We
might have insisted that the baryon asymmetry was the initial
condition of the universe.  However, this is not possible because we
will accept the inflationary paradigm.  We will come back to this
point later.  Therefore, the asymmetry needs to be explained. In fact,
having accepted two right-handed neutrinos, we can let them produce
the baryon asymmetry via leptogenesis
\cite{Fukugita:1986hr,Frampton:2002qc,Endoh:2002wm}.  This is possible
only for Majorana neutrinos with seesaw mechanism without additional
degrees of freedom, unlike leptogenesis with Dirac neutrinos
\cite{Dick:1999je}.  Therefore, we do not have a choice: the neutrinos
are Majorana, and the decays of right-handed neutrinos in the early
universe, coupled with the electroweak anomaly, is responsible for
creating the baryon asymmetry.  The NMSM Lagrangian, hence, must also
include
\begin{equation}
  \label{eq:N}
  {\cal L}_N =
  \bar{N}_\alpha i{\not\!\partial} N_\alpha -
  \left( \frac{ M_\alpha}{2} N_\alpha N_\alpha
    + h_\nu^{\alpha i} N_\alpha L_i \tilde{H} + c.c. \right).
\end{equation}

Because the left-handed neutrino Majorana mass matrix has rank two,
there is one massless state.  The other two neutrino masses can be
determined from the solar and atmospheric neutrino data, and there is
only one Majorana phase.  In the basis where the charged-lepton and
right-handed-neutrino mass matrices are real and diagonal, there are
eleven real parameters in Eq.~(\ref{eq:N}), after rephasing of three
lepton doublets.  Since there are only seven real parameters for light
neutrinos, two masses, three mixing angles, one Dirac and one Majorana
phase, we have enough parameters to accommodate the current data. In
order to produce the observed baryon asymmetry via leptogenesis, the
lighter right-handed neutrino should be heavier than $10^{10}$ GeV to
have enough CP asymmetry~\cite{Endoh:2002wm, Barger:2003gt}.

%% Because the left-handed neutrino Majorana mass matrix is rank two
%% due to two right-handed neutrinos,
%% there is one massless left-handed neutrino. Then,
%% the active neutrino masses can be determined from
%% the solar neutrino and atmospheric neutrino
%%  experiments, and there is only one Majorana phase.
%% In the basis where the lepton and right-handed neutrino mass
%% matrices are real and diagonal,
%% we can reconstruct the neutrino Dirac mass matrix
%% from one of its six entries, the right-handed and left-handed
%% neutrino masses,
%% the neutrino mixing angles, a Dirac phase and a Majorana
%% phase~\cite{Barger:2003gt}.
%% Thus, there is no problem to explain the active neutrino masses
%% and bilarge mixings from reconstruction. In order to produce the observed
%% baryon asymmetry via leptogenesis,  the lightest
%% right-handed neutrino should be heavier than $10^{10}$ GeV
%% so that we have enough CP
%% asymmetry~\cite{Endoh:2002wm, Barger:2003gt}.
%% Interestingly, if we have one or two constraints on
%% the neutrino Dirac mass matrix, for example, texture
%% zero or horizontal equality, we can have the correlation
%% between the $CP$-violating phases in the leptogenesis and
%% low-energy neutrino experiments (neutrino oscillation
%% and $0\nu \beta \beta$ decay), which might provide a way
%% to test leptogenesis~\cite{Endoh:2002wm, Barger:2003gt}.

Finally, nearly scale-invariant, adiabatic, and Gaussian density
fluctuations need to be generated in order to explain the observed
structure, velocity field, and cosmic microwave background anisotropy.
We adopt inflation for this purpose.  We do not see any candidate
scalar field to drive inflation in the MSM nor among the new particles
introduced above.  Therefore, we have to introduce at least another
degree of freedom.  The minimal new particle content is again a real
KG field, and its most general renormalizable Lagrangian is
\begin{equation}
  \label{eq:inflaton}
  {\cal L}_\varphi =
  \frac{1}{2} \partial_\mu \varphi \partial^\mu \varphi
  - \frac{1}{2} m^2 \varphi^2 - \frac{\mu}{3!} \varphi^3
  - \frac{\kappa}{4!} \varphi^4.
\end{equation}
Here, the possible linear term has been absorbed by a shift.  This
potential can drive inflation, {\it e.g.}\/, if the field starts with
a trans-Planckian amplitude; this is nothing but the chaotic inflation
model \cite{Linde:gd}.  Current data prefer the quadratic term to
drive inflation \cite{Peiris:2003ff,Seljak:2004xh} with $m \simeq 1.8
\times 10^{13}$~GeV \cite{Ellis:2003sq}, while $\mu \lesssim
10^{6}$~GeV and $\kappa \lesssim 10^{-14}$.%
%% \footnote{Other choices of parameters may also drive
%%   inflation successfully.  For example, when $m^2\approx 0$, the field
%%   can start rolling down from the origin due to the cubic term, while
%%   $m^2 < 0$ can drive inflation from the origin as well.  For our
%%   purposes, it is enough that there is one parameter set that works.}.
\footnote{It may well be possible to achieve successful inflation also
  with small field amplitudes (small-field models), but many existing
  models require more than one degree of freedom; we do not pursue
  this interesting possibility further in this letter.  }

The only possible renormalizable couplings of the inflaton to
other fields in the NMSM allowed by symmetries are
\begin{eqnarray}
  \label{eq:reheat}
  V_{RH} &=& \mu_1 \varphi |H|^2 + \mu_2 \varphi S^2
  + \kappa_H \varphi^2 |H|^2 + \kappa_S \varphi^2 S^2 \nonumber \\
  & & + (y_{N}^{\alpha\beta} \varphi N_\alpha N_\beta + c.c.).
\end{eqnarray}
Reheating after inflation can take place by couplings $\mu_1$,
$\mu_2$, or $y_N^{\alpha\beta}$.  For thermal leptogenesis to take
place, the reheating temperature must be higher than the mass of the
lighter right-handed neutrino, say $10^{10}$~GeV, requiring either
$\mu_{1,2} \gtrsim 10^9$~GeV or $y_{N}^{\alpha\beta} \gtrsim 10^{-4}$;
they do not spoil the flatness of the inflaton potential if
$\kappa_{H,S} \lesssim 10^{-6}$.  Moreover, $y_N^{\alpha\beta}$ lets
the inflaton decay directly to the right-handed neutrinos, whose
subsequent decay can produce the asymmetry~\cite{Lazarides:wu,
  Asaka:1999jb}, allowing for even smaller couplings.  This is a
non-trivial cross check that the inflation and the leptogenesis are
consistent within our model.

Let us come back to the question if the baryogenesis is necessary.
Even if we accept the inflationary paradigm, one may still hope that a
large initial baryon asymmetry before the inflation may be retained to
account for the observed value.  We can exclude this possibility on
purely empirical grounds.  Even if we set aside the desire to explain
the horizon and flatness puzzles, which are after all aesthetic issues
which we disregard in this letter, we have just accepted inflation as
the source of nearly scale-invariant density fluctuations to account
for the cosmic microwave background anisotropies, large scale
structures, and eventually galaxy formation.  Therefore we need the
$e$-folding of the inflation to be larger than the logarithm of the
ratio of the cosmological scale to the galactic scale, conservatively
$N\gtrsim \ln (10\mbox{Gpc}/10\mbox{kpc}) = 14$.  On the other hand,
the large intial baryon asymmetry before the inflation can only be in
the form of a Fermi-degenerate gas.  Its energy density $\rho_B \simeq
\mu_F^4$, where $\mu_F$ is the Fermi momentum, behaves as radiation.
In order for the inflation to start, the energy density of the
Fermi-degenerate gas must be less than that of the inflaton
$\rho_\phi$.  Assuming that they were approximately the same, the
energy density of the baryon gas is suppressed by $\mu_F^4/\rho_\phi
\simeq e^{-4N}$ at the end of the inflation.  Reheating will further
dilute the baryon asymmetry and hence we conservatively assume that
the reheating was instantaneous.  Then the maximum baryon asymmetry
one can obtain is $\eta \simeq \mu_F^3/\rho_\phi^{3/4} \simeq e^{-3N}
\lesssim 10^{-18}$, insufficient to explain the observed asymmetry of
$\eta \simeq 10^{-10}$.  Therefore, baryon asymmetry cannot be
explained by the initial condition based on purely empirical arguments
once inflation is accepted as the source of the density fluctuations.

It is remarkable that the MSM Lagrangian Eq.~(\ref{eq:MSM}),
supplemented by the most general renormalizable Lagrangian in
Eqs.~(\ref{eq:S}, \ref{eq:Lambda}, \ref{eq:N}, \ref{eq:inflaton},
\ref{eq:reheat}) for two right-handed neutrinos $N_\alpha$, one $Z_2$
odd real scalar $S$, and another real scalar $\varphi$,
\begin{equation}
  \label{eq:4}
  {\cal L}_{NMSM} = {\cal L}_{MSM} + {\cal L}_S + {\cal L}_\Lambda
  + {\cal L}_N + {\cal L}_\varphi - V_{RH},
\end{equation}
explains everything we currently know about our universe.

This model is supposed to describe all known physics including
classical gravity.  Note that quantum gravity effects have not
empirically been observed and hence are beyond the scope of the
model, but we expect them to be there.  Thus we assume there is no
new physics beyond the NMSM up to the Planck scale.  All higher
dimension operators from the cut-off scale are suppressed by the
Planck scale. Hence it is free from excessive flavor-changing
effects, CP violation, too-rapid proton decay, and problems with
electroweak precision data.

%% What do we know about the parameters of the NMSM?  For the MSM to
%% be valid up to the Planck scale, various authors have studied
%% constraints from the instability and triviality of the Higgs
%% potential (see, {\it e.g.}\/, \cite{Lindner:1985uk}).  We do the
%% same for the NMSM.  At the one-loop level, the gauge coupling
%% constants run according to the following renormalization-group
%% equations,
%% \begin{eqnarray}
%%   \label{eq:7}
%%   (4\pi)^2 \frac{d g^{\prime}}{dt} &=& \frac{41}{6}
%%   g^{\prime 3}, \\
%%   (4\pi)^2 \frac{d g}{dt} &=& - \frac{19}{6} g^3, \\
%%   (4\pi)^2 \frac{d g_s}{dt} &=& - 7 g_s^3.
%% \end{eqnarray}
%% Here and below, $t = \log \mu$.  The only Yukawa coupling important
%% for our purposes is the top Yukawa coupling $y$ which runs as
%% \begin{equation}
%%   \label{eq:1}
%%   (4\pi)^2\frac{dy}{dt} = y \left(
%%     \frac{9}{2} y^2 - \frac{17}{12} g^{\prime 2} - \frac{9}{4} g^2
%%     - 8 g_s^2 \right).
%% \end{equation}
%% They are the same as in the MSM.  The couplings in the scalar sector
%% run as

\begin{figure}
  \centering
  \includegraphics[width=\columnwidth]{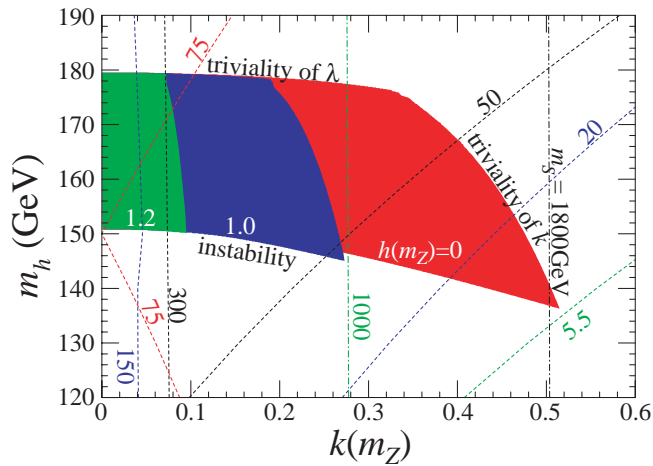}
  \caption{The region of the NMSM parameter space $(k(m_Z), m_h)$ that
    satisfies the stability and triviality bounds, for $h(m_Z)=0$,
    1.0, and 1.2.  Also the preferred values from the cosmic abundance
    $\Omega_S h^2 = 0.11$ are shown for various $m_S$.  We used
    $y(m_Z)=1.0$.}
  \label{fig:params}
\end{figure}

%What do we know about the parameters of the NMSM?  
Now we come to another non-trivial consistency check of the model,
that is the addition of the scalar $S$ does not conflict with
empirical requirements.  For the MSM to be valid up to the Planck
scale, various authors have studied constraints from the instability
and triviality of the Higgs potential (see, {\it e.g.}\/,
\cite{Lindner:1985uk}).  We do the same for the NMSM.  At one-loop
level, the gauge coupling constants and the top Yukawa coupling $y$
run the same way as in the MSM.  The couplings in the scalar sector
run as
\begin{eqnarray}
  \label{eq:2}
  (4\pi)^2 \frac{d\lambda}{dt}  &=&
    12 \lambda^2 + 12 \lambda y^2 - 12 y^4
    - 3\lambda(g^{\prime 2}+3g^2)  \nonumber \\
  & &
    + \frac{3}{4} \left[2g^4 + (g^{\prime 2}+g^2)^2\right]
    + k^2 , \\
  (4\pi)^2 \frac{dk}{dt} &=& k\left[4 k
    + 6 \lambda + h + 6 y^2
    - \frac{3}{2} (g^{\prime 2}+3g^2)\right], \\
  (4\pi)^2 \frac{dh}{dt} &=& 3 h^2 + 12 k^2,
\end{eqnarray}
with $t = \log \mu$.  We require that none of the couplings be driven
negative below the Planck scale (stability bound) and stay below 10
(triviality bound).  The region of $(m_h, k(m_Z))$ is shown in
Fig.~\ref{fig:params} for three values of $h(m_Z) = 0, 1.0, 1.2$.  The
region disappears when $h(m_Z) \gtrsim 1.3$.  The Higgs boson is
predicted to be light, at most 180~GeV, while heavier than 130~GeV.
This range is in complete accordance with the precision electroweak
fits $m_h \lesssim 200$~GeV \cite{LEPEWWG:2003ih}, while beyond the
LEP-II reach \cite{Barate:2003sz} and is not probed experimentally
yet.
%% On the other hand, discovery of a Higgs boson outside this range
%% would require new physics beyond the NMSM.

The Dark Matter annihilation cross section is proportional to $k^2$
and depends on $m_S$ and $m_h$ \cite{McDonald:1993ex}.  We have
improved the abundance calculation using HDECAY \cite{Djouadi:1997yw}
and included the $s$-channel Higgs exchange diagram in $SS \rightarrow
hh$, absent in \cite{McDonald:1993ex} even though it is not
qualitatively important.  Preferred values of $(k(m_Z), m_h)$ are
shown for $\Omega_S h^2 = (\Omega_m-\Omega_b) h^2 = 0.11$ as curves in
Fig.~\ref{fig:params} for various $m_S$.  Note that $m_S = 75$~GeV
allows for annihilation through Higgs pole and has a special behavior.
To be consistent with the triviality and stability bounds, we find
$m_S \simeq 5.5$~GeV--1.8~TeV.

\begin{figure}
  \centering
  \includegraphics[width=\columnwidth]{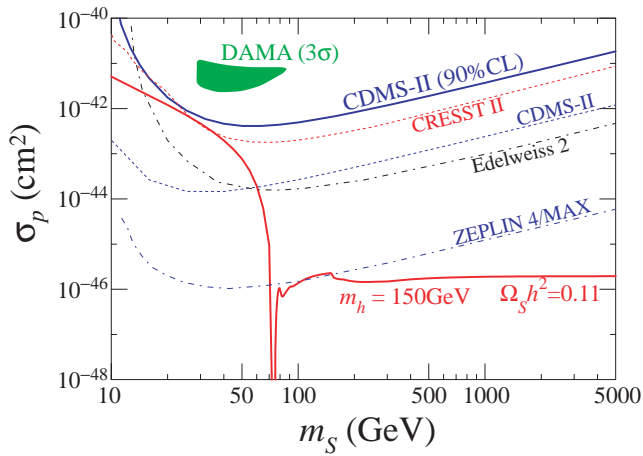}
  \caption{The elastic scattering cross section of Dark Matter
    from nucleons in NMSM, as a function of the Dark Matter particle
    mass $m_S$ for $m_h=150$~GeV.  Note that the region $m_S \gtrsim
    1.8$ TeV is disallowed by the triviality bound on $k$.  Also shown
    are the experimental bounds from CDMS-II \cite{:2004fq} and DAMA
    \cite{Bernabei:2003za}, as well as improved sensitivities expected
    in the future \cite{dmtools}. }
  \label{fig:sigmap}
\end{figure}

Now we have demonstrated that all new elements we have added to the
MSM do not cause any tensions among themselves nor with the empirical
constraints.  The new scalar we added at the TeV-scale is consistent
with the electroweak data even after we imposed the triviality and
stability bounds, while it can give the required cosmological density
without conflicting the direct search limits.  It does not induce any
flavor-changing effects or new CP violation that typically haunt
models with new degrees of freedom at the TeV scale.  The inflation
model we adopted can successfully reheat to a high-enough temperature
to account for leptogenesis for parameters consistent with neutrino
oscillation data, while the required coupling for the reheating does
not spoil the required flatness of the inflaton potential.  We also
pointed out that inflation, even with a conservative requirement on
the $e$-folding based on purely empirical grounds, actually requires
baryogenesis.

Are there new observable consequences of the NMSM?  The Higgs boson
may decay invisibly $h \rightarrow SS$ \cite{Burgess:2000yq}.  It will
be subject to search at the LHC via $W$-boson fusion, or more
promisingly at a Linear Collider.  If the singlet is heavier than
$m_h/2$, the search at collider experiments becomes exceedingly
difficult.  One possibility is the $W$-boson fusion processes $qq
\rightarrow qq SS+g$ or $qqSS+\gamma$, where forward jets are tagged,
large missing $p_T$ is seen, together with additional isolated photon
or jet.  It may not cover the entire range up to 1.8~TeV.  The
scattering of $S$ on nuclei is dominated by the Higgs boson exchange,
as worked out in \cite{McDonald:1993ex,Burgess:2000yq}.  The
prediction for $m_h = 150$~GeV is shown in Fig.~\ref{fig:sigmap}; it
is clear that the model is consistent with the current limit from
CDMS-II \cite{:2004fq}.  It cannot explain, however, the controversial
data from DAMA \cite{Bernabei:2003za}.  Because the Higgs boson is
light thanks to the triviality bound, the scattering cross section is
promising for the underground Dark Matter searches for $m_S \lesssim
m_h/2$.

The spectrum index of the $\varphi^2$ chaotic inflation model is
predicted to be 0.96.  This may be confirmed in improved
cosmic-microwave background anisotropy data, with more years of
WMAP and Planck.  The tensor-to-scalar ratio is 0.16
\cite{Ellis:2003sq}, again within the reach of near future
observations.  For other inflationary scenarios, predictions vary.
The equation of state of Dark Energy is predicted to be exactly
$w=-1$.

Neutrinos are Majorana fermions and hence we expect neutrinoless
double beta decay at some level.  Because one of the neutrino masses
exactly vanishes (ignoring tiny Planck suppressed effects), the signal
in the near-future experiments is possible only for the inverted
hierarchy \cite{Murayama:2003ci}.

%% and null result from KATRIN tritium beta decay 
%% would imply the
%% inverted hierarchy \cite{Bahcall:2004ip}.  If the hierarchy is
%% normal, however, it would be difficult to observe neutrinoless
%% double beta decay.

Here we list a few future observations that could rule the NMSM
incomplete.  Obviously, discovering any particles at the
electroweak scale other than $h$ and $S$ at a collider will require an
extension of the model.  A Higgs mass inconsistent with the bounds in
Fig.~\ref{fig:params} will also be a smoking gun for additional
physics.  Confirmation of the DAMA signal would require a different
Dark Matter candidate.  Signals of some rare decays, such as $\mu \to
e \gamma$, would require extra flavor-changing effects.  Observation
of new sources of CP violation beyond the CKM and MNS phases is
another avenue, {\it e.g.}\/, an electron electric dipole moment or a
discrepancy in $\sin 2 \beta$ between $B \to \phi K_S$ and $\psi K_S$
modes.  As for the neutrino sector, a confirmation of the LSND results
by the Mini-BooNE experiment would require new degrees of freedom
beyond the NMSM.  Positive signal for neutrino mass at KATRIN would
require masses for all three neutrinos.  A future observation by a
satellite experiment, such as Planck, of $\Omega_{\rm tot}$ deviating
from unity or of non-Gaussianity of the density fluctuations could
rule out the one-field inflationary scenario of the NMSM.  Finally,
detection of proton decay in any of the current or foreseeable future
experiments cannot be explained in the NMSM.

It needs to be mentioned that the NMSM does require an extreme degree
of fine-tuning.  The cosmological constant represents a tuning with an
accuracy of $10^{-120}$.  The hierarchy between the electroweak and
the Planck scales should also be fine-tuned at the level of
$10^{-32}$. Fermion mass hierarchies and mixings are not explained.
The QCD vacuum angle is simply chosen to be $\theta \lesssim
10^{-10}$.  The $Z_2$ symmetry on the singlet is imposed by hand.
%% The inflaton potential is fine-tuned to be flat enough
%% for its purpose. 
The parameters in the inflation potential are chosen to be small.
Nonetheless, the model is empirically successful
in describing everything we know about fundamental physics, and
needs to be taken seriously.  Any new physics beyond the NMSM that
may address the aesthetic issues mentioned here should not spoil
the success of the NMSM.

Here, we list some possible directions for going beyond the scope of
the present work.  The triviality and stability bounds can be improved
to two-loop level.  Feasibility of collider searches for $S$ with $m_S
> m_h/2$ needs further analysis.  For this mass region, indirect Dark
Matter searches are of great interest, since both collider and direct
Dark Matter searches are challenging.  It would require a detailed
Monte Carlo study of the annihilation products in the Sun.  A lighter
$S$ can be seen in the invisible decay of the Higgs boson at a Linear
Collider, while its mass measurement would require an off-shell Higgs
process which needs to be investigated.  Other possibilities for the
one-field inflationary scenario may warrant further study.

In summary, we have presented the New Minimal Standard Model of
particle physics and cosmology that incorporates Dark Matter, Dark
Energy, neutrino masses and mixings, baryon asymmetry, and nearly
scale-invariant Gaussian density fluctuations, based on the
principle of minimal particle content and the most general
renormalizable Lagrangian.  Remarkably, it requires only six new
degrees of freedom. Any model of physics beyond the Minimal
Standard Model should be judged against the empirical success of
this model.

\acknowledgments

RK, the Marvin L.~Goldberger Member at the Institute for Advanced
Study (IAS), gratefully acknowledges support for his research.  This
work was supported by the IAS, funds for Natural Sciences, as well as
in part by the DOE under contracts DE-FG02-90ER40542 and
DE-AC03-76SF00098, and in part by NSF grants PHY-0098840 and
PHY-0070928.

%%%%%%%%%%%%%%%%%%%%%%%%%%%%%%%%%%%%%%%%%%%%%%%%%%%%%%%%%%%%%%%%%%%%%%%%%%%%

\end{document}